\documentclass[preprint,12pt]{elsarticle}

\usepackage{amssymb}
\usepackage{amsmath}
\usepackage{multirow}

\journal{Journal of Information Security and Applications}

\begin{document}

\begin{frontmatter}



\title{A novel TLS-based Fingerprinting approach that combines feature expansion and similarity mapping}


\author{Amanda Thomson, Leandros Maglaras, Naghmeh Moradpoor} 

\affiliation{organization={Edinburgh Napier University},
            addressline={10 Colinton Road.}, 
            city={Edinburgh},
            postcode={EH10 5DT}, 
            country={UK}}

\begin{abstract}
Malicious domains are part of the landscape of the internet but are becoming more prevalent and more dangerous to both companies and individuals. They can be hosted on variety of technologies and serve an array of content, ranging from Malware, command and control, and complex Phishing sites that are designed to deceive and expose. Tracking, blocking and detecting such domains is complex, and very often involves complex allow or deny list management or SIEM integration with open-source TLS fingerprinting techniques. Many fingerprint techniques such as JARM and JA3 are used by threat hunters to determine domain classification, but with the increase in TLS similarity, particularly in CDNs, they are becoming less useful. The aim of this paper is to adapt and evolve open-source TLS fingerprinting techniques with increased features to enhance granularity, and to produce a similarity mapping system that enables the tracking and detection of previously unknown malicious domains.  This is done by enriching TLS fingerprints with HTTP header data and producing a fine grain similarity visualisation that represented high dimensional data using MinHash and local sensitivity hashing. Influence was taken from the Chemistry domain, where the problem of high dimensional similarity in chemical fingerprints is often encountered. An enriched fingerprint was produced which was then visualised across three separate datasets. The results were analysed and evaluated, with 67 previously unknown malicious domains being detected based on their similarity to known malicious domains and nothing else. The similarity mapping technique produced demonstrates definite promise in the arena of early detection of Malware and Phishing domains. 

\end{abstract}

\begin{graphicalabstract}
\begin{figure}[htb!]
     \centering
     \includegraphics[width=.8\linewidth]{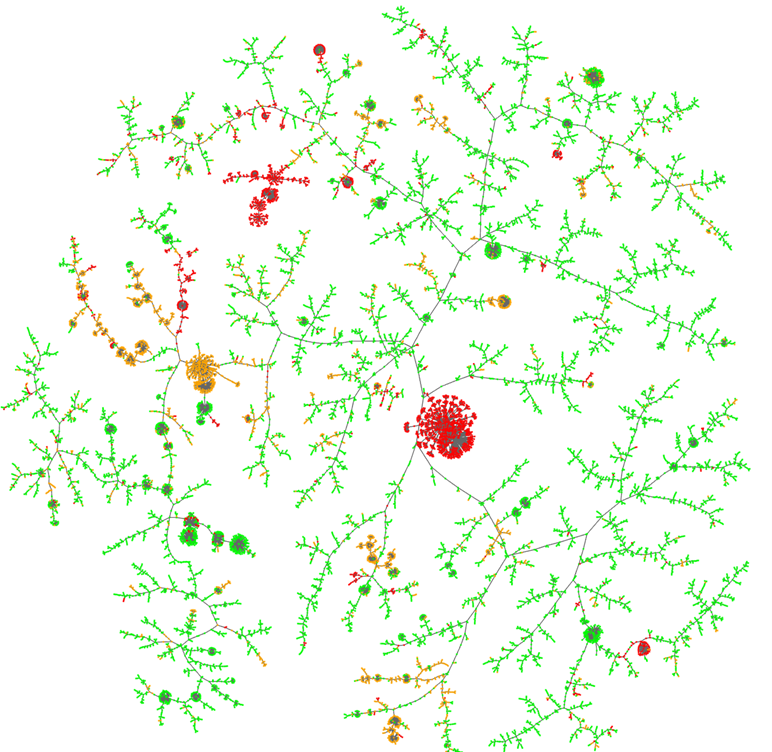}
     \caption{Graph displaying TLS features enriched with HTTP header data. Known good domains are coloured green, known bad red, and unknown orange.}
     \label{fig:fearun_1}
 \end{figure}
\end{graphicalabstract}

\begin{highlights}
\item The article conducts a critical literature review evaluating active scanning techniques that can be used to generate server fingerprints.
\item Presents the design and development of an active scanning fingerprint.
\item The proposed method increases the granularity of current techniques, improving the ability to detect malicious domains hosted on CDNs where feature similarity is high.
\item Enhance the applicability and security of fingerprinting by introducing a suitable similarity mapping approach.
\item Critically evaluate the results and findings form the practical experiments.
\end{highlights}


\end{frontmatter}



\section{Introduction}
\label{sec1}

There is an increasing threat of malicious domains to users and networks, whether that’s Phishing domains that are designed to harvest user information and credentials or malicious domains that host command and control servers. There are several open-source techniques and tools designed to assist with the classification of domains, but many of them are reactionary and rely on reporting from third parties, tools or systems. One key arsenal in the weapon of classification however is active TLS fingerprinting. The approach uses scanning to conduct exhaustive probes on domains and amalgamates the data returned from the Client Hello’s to determine the configuration of the server from those features. 

These TLS features, whilst appearing benign can help form a picture of underlying technologies and libraries, such as the version of OpenSSL. Given the fingerprint of a malicious domain, it’s possible to pivot to other domains with the same fingerprint and identify malware with the same technology ad configurations. Given the high level of automation in common malware, ransomware and phishing tools, it is reasonable to assume that many malicious domains will share similar configurations and deployments. 

Fingerprinting in this manner is used widely in tooling such as Shodan and Censys, and whilst useful for threat hunting and pivoting, it is not fool proof. If the TLS features included in the fingerprint are too wide, then cardinailty will be too high and result in high levels of benign domains sharing fingerprints with malicious domains. Many of the methods currently in use also rely on hash-based approaches, meaning it’s difficult to understand how similar domains are, as small differences in configurations produce radically different hashes. 

This paper aims to determine if the hypothesis that malicious domains can be classified more easily based upon their similarity to other domains is correct, rather than exclusively relying on hash-based approaches. It aims to examine the current state of the art for active scanning-based fingerprinting and determine if increased granularity of server features can aid in detection of malicious domains even when they are hosted on CDNs – which greatly narrows the range of TLS features.

The main contributions of the current article are:

\begin{itemize}
    \item  Conducts a critical literature review evaluating active scanning techniques that can be used to generate server fingerprints.
    \item  Applies the information gathered in the literature review to the design and development of an active scanning fingerprint that increases the granularity of current techniques, improving the ability to detect malicious domains hosted on CDNs where feature similarity is high.
    \item  Enhance the applicability and security of fingerprinting by introducing a suitable similarity mapping approach, making it difficult to subvert hash-based fingerprints with minuscule adjustments or manipulations. 
    \item  Critically evaluate the results and findings form the practical experiments.
\end{itemize}

The paper is structured as follows: Section 2 presents an in-depth literature review to determine the current state of the research and what gaps can be identified. Section 3 presents the methodology for the practical experiments and examines any decisions made. Section 4 presents the results of the practical in-depth and extensive evaluations, including a comparison to others' research. Finally, section 5 summarises the work done and suggests future research and avenues for improvement. 
\cite{}

\section{Related Work}

The first version of the TLS protocol was published in 1999 by the Internet Engineering Task Force (IETF) and has been under constant review and development since, with the latest version, TLS 1.3 \cite{rescorla2018transport} being released in 2018. Despite such a lengthy period of adoption, it remains in the Open Web Application Security Projects (OWASP) top ten and as recent as 2021 cryptographic failures were promoted to the number two slot \cite{warburton20212021}. According to the OSWAP report, underlying issues can range from a simple lack of understanding of the TLS mechanisms resulting in weak cipher suite choices, to invalid certificate chains and depreciated TLS versions. 

TLS 1.3 is increasingly being adopted for its enhanced security and privacy features, with 63\% of the global top 1 million web sites now supporting it \cite{warburton20212021}. Despite the modernity of the version however, there remained gaps in the privacy aspect of the protocol, and in 2022 the TLS working group released a further amendment to the protocol \cite{rescorla2018transport} to introduce the Encrypted Client Hello (ECH). The introduction of ECH has far reaching implications on some TLS fingerprinting methods that have traditionally relied on metadata contained with the unencrypted client hello.

As this paper is focused on the future ability to detect and fingerprint Malware and C2 domains under TLS, the background and exposition work will be focused on the latest released version 1.3 \cite{rescorla2018transport} with previous versions only referenced for historical clarity. 

\subsection{Active Fingerprinting}
Although the goal of TLS is to improve security and privacy, a side effect is that it hampers the ability to differentiate malicious traffic from normal traffic. As the volume of companies utilising security operations centres (SOCs) increases, the ability to classify traffic correctly is a valuable tool. Oh et al.\cite{oh_survey_2022} raise the issue that Network Traffic Analysis (NTA) is becoming increasingly more difficult due to 80\% of internet traffic now being under the HTTPS protocol. They highlight one of the motivating factors in this growth as being the ease of access to certificate providers, specifically Let’s Encrypt, but also the drive by large web browser technologies such as Mozilla and CDN providers like CloudFlare. Shamsimukhametov et al.\cite{shamsimukhametov_is_2022} raise similar concerns regarding the uptake in malware utilising TLS but focus on the requirement for privacy on the internet.

Active fingerprinting is the technique of actively identifying malicious domains or web servers and can be useful in several scenarios including threat hunting, infrastructure hunting and for detecting the use of CDNs \cite{sosnowski_active_2023}. These methods involve making active, but benign connections to a domain in order to gather enough information to make an accurate classification as to the intent of the server. The most common technique currently in use is known as JARM \cite{lindeman_easily_2020}. A JARM is a fuzzy SHA256 hash of the resultant selected set of metadata extracted from ten TLS connections. Each of the ten connections uses a different ClientHello, forcing the server to respond with a range of results that enables profiling of the server. The server behaviours \cite{sosnowski_active_2023} are considered the totality of the servers capabilities, made up from the configuration of the TLS which can influence the handshake procedure. This indicates that servers with identical JARMs will have very similar software configurations. 

In comparison to JARM which creates a fingerprint based on unencrypted TLS metadata, Sosnowski et al. \cite{sosnowski_active_2023} created Active TLS Stack Fingerprinting. In contrast to the JARM methodology, the tooling enables users to configure not only the number of ClientHello’s sent but the configuration of the ClientHello’s themselves. Another key difference is the full completion of the TLS connection during the active scan, enabling the tooling to collect metadata that would otherwise be encrypted in the new ECH. The feature selection used to build the fingerprint consists of the TLS version, cipher suites, any received alerts and extensions data. 

Although Sosnowski et al. \cite{sosnowski_active_2023} conclude that their tooling can differentiate 55\% more server behaviours than the JARM approach, the authors still had to incorporate additional metadata in the form of HTTP headers captured during the scan to further improve the ability to detect command and control (C2) servers. Papadogiannaki and Ioannidis,\cite{papadogiannaki_pump_2023} also highlighted some weaknesses with the JARM technique, similar to the issues raised by \cite{matousek_reliability_2021}. They stated that the volume of collisions with benign domains can mean diminishing returns if the database is not kept up to date. This was reflected in their results, with the overlap of malicious fingerprints with benign being only 135 in 2021 versus 40\% in 2022. 

\subsection{The CDN Problem}
Content delivery networks are geographically distributed nodes whose purpose is to improve the speed and availability of content by moving it closer to the end user. They are increasingly being adopted to help with distributed denial of service (DDOS) attacks and some CDN vendors have been instrumental in pioneering the early adoptions of technologies like ECH, with Cloudflare enabling the technology as recently as 2023 \cite{van_der_mandele_encrypted_2023}. 

The increase adoption of CDNs has naturally had an impact on the statistics of TLS usage. According to the http archive Web Almanac 87\% of sites using TLS now use the latest version compared to only 42\% of non CDN origin sites \cite{bhandari_2022_2022}. This is driven by the fact that CDNs handle the TLS termination at the edge, and create a second connection back to the origin. The fact that the CDN handles the termination means that TLS configurations are often no longer done by the origin server and as the CDN can be configured to handle updates and configuration changes, more uniformity across TLS profiles is seen.
Siby et al. \cite{siby_evaluating_2023} discuss some of the issues faced when fingerprinting websites hosted behind CDNs. They highlight the increased usage of CDNS, with 44\% of the top 1 million sites now utilising them, and how the co-hosting of multiple domains behind a singular IP, with the same TLS configuration creates a natural anonymity for the origin server. Sosnowski et al. \cite{sosnowski_active_2023} help confirm the principal difficulties behind CDN identification focusing just on TLS metadata during their development of the Active TLS Stack fingerprinting tool. Their attempt to identify CDN deployments outside of a CDN’s AS clearly highlighted the similarity between TLS fingerprints, with their research allocating specific fingerprints to several large CDN providers. Although the ranges of fingerprints were small in some cases, such as Alibaba being assigned only 1 fingerprint, Cloudflare was allocated 801 fingerprints, more than would be expected for a provider deploying TLS-enabled servers at scale. 

\subsection{TLS Fingerprint Enrichment}
A large volume of the research into TLS fingerprinting suggests that it’s a viable tool for clustering nodes with similar configurations, but is not granular enough on its own to enable true classification. As the role out of TLS 1.3 and ECH progresses along with the uptake in use of CDN’s, the variation’s seen across the TLS stack is decreasing, making TLS fingerprinting less viable as a fine-grained method for malicious domain and infrastructure hunting. 

Sosnowski et al. \cite{sosnowski_active_2023} included the HTTP 'Server' key in their initial scans to help detect CDN server deployments outside of a CDNs designated AS. The additional HTTP metadata when combined with TLS, achieved a maximum precision of 97\%, but the authors were clear that HTTP header analysis alone was not ideal as a classifier. In contrast to this, Tang et al. \cite{tang_hslf_2021} developed HSLF, a local sensitive hashing algorithm that sequenced all header fields and measured similarities between HTTP sessions. The authors chose to weight specific headers that reflected application specific details before applying a random forest machine learning algorithm that resulted in a accuracy rating on 0.96 when classifying application specific traffic.

HTTP Headers have also been used to track and pivot between malicious domains, McGahagan et al. \cite{mcgahagan_comprehensive_2019} analysed numerous headers from benign and malicious web-servers and used eight machine learning models to demonstrate the feasibility and effectiveness of just using headers in isolation. The authors chose to invest heavily in data cleaning and the validation process, ensuring that custom headers and misspellings were not considered. This approach was in direct conflict to the one taken by Al-Hakimi and Bax, \cite{al-hakimi_hunting_2021} who used headers for hunting malicious infrastructure. Their research considered misspellings and unusual sequencing to be valuable anomalies that could aid in the identification of unique command and control servers. The importance of header sequences was echoed by Bortolameotti et al. \cite{bortolameotti_headprint_2020} who produced HeadPrint to identify malicious communications in passive traffic. Their fingerprint technique relied on two orthogonal header characteristics that enabled applications to be distinguished – their order and their associated values. A range of machine learning models were evaluated using the fingerprinting technique and the accuracy ranged from 90.74\% to 95.44\%.  

Although research into the domain of active HTTP header fingerprint is limited, the current research does indicate that, at least on an application level, HTTP headers can be effective in fingerprinting \cite{al-hakimi_hunting_2021} \cite{bortolameotti_headprint_2020} \cite{tang_hslf_2021}, and is a useful candidate for enrichment of TLS fingerprints. 

\section{Methodology}
It's clear from the current research that as technology progresses, TLS fingerprints are becoming less granular, which naturally limit's their effectiveness. This section details the implementation of the proposed active scanning fingerprint enrichment approach, beginning with the data acquisition, sensitisation and preparation, followed by the similarity mapping and visualisation of the final enriched fingerprint. 

The tool chosen for the scanning was the ActiveTLS stack fingerprinting tool \cite{sosnowski_efactls_2024-1}. The tools offers greater granularity of features when compared to JARM, handles connections that make use  of the encrypted client hello and produces output that is not obfuscated, providing a greater scope for onward processing. 

\subsection{Data Acquisition}

In preparation for the scanning, datasets of known good, known bad and unknown domains were acquired from several sources. All good domains were taken from the Tranco list \cite{le_pochat_tranco_2019}, a highly regarded, robust list of good domains that is considered resistant to manipulation. Known active malicious, and unknown domains were taken from a range of sources which can be seen in Table \ref{tab:Table_1}. Unknown domains were included in the test scenario's to aid in process evaluation, and are domains that have neither been flagged as known good or known malicious. As an average of 20\% of all newly registered domains are usually established as being malicious \cite{tilborghs_flagging_2022} they are useful candidates for demonstrating effectiveness of the similarity mapping. 
\begin{table}[htb]
    \centering
    \begin{tabular}{cccc} 
    \hline
     Data Set&Location&Category\\
    \hline
    \hline
        Tranco LJNY4 & https://tranco-list.eu & Good\\
        UrlHaus	& https://urlhaus.abuse.ch & Bad\\	
        Hunt.io	& https://hunt.io & Bad \\ 
        Cert.pl	& https://hole.cert.pl & Bad \\ 
        OpenPhish & https://openphish.com & Bad\\ 
        Shreshtait & https://shreshtait.com & Unknown \\ 
    \end{tabular}
    \caption{Table of All Datasets Used For Active Scanning.}
    \label{tab:Table_1} 
\end{table}

The initial steps involved resolving the domains to the their respective IP address. To enable this, the stub DNS resolver MassDNS \cite{Blechschmidt} was used. The output {domain:ip} format was then appended with the 10 curated ClientHellos used by the ActiveTLS tool to ensure parity of results across the datasets. The order of of domains was randomised to ensure the load on servers was distributed across the 10 scans and the TLS extended output was enabled. Importantly, the HTTP scan functionality was not enabled within the ActivetTLS too itself, but instead the headers were gathered in a a post processing step. This ensured the order of the headers could be maintained and prevented having to deconflict headers across multiple scans.  

\subsection{Post Processing}
The fingerprint produced by the ActiveTLS tool can be seen in Figure \ref{fig:Original_fingerprint}. To ensure the maximum flexibility for further enrichment and similarity mapping, the fingerprint itself was broken down into its constituent element’s version, ciphers, extensions, encrypted extensions and certificate extensions. The raw fingerprint itself was kept, but SHA256 encoded to make readability easier and make manual similarity identification simpler. 

\begin{figure}[h]
\centering
771\_1302\_43.AwQ-51.23\_0.-16.AAMCaDI\_\_43.AwQ-51.23\_-|771\_c030\_0.-1.AQ-16.AAkIaHR0cC8xLjE\_\_\_\_|771\_c02f\_65281.-0.-11.AwABAg-35.-16.AAMCaDI\_\_\_\_|771\_1301\_43.AwQ-51.29\_0.-10.AAQAFwAd\_\_\_-|\_\_\_\_\_\_<40|771\_1302\_43.AwQ-51.23\_0.-16.AAMCaDI\_\_43.AwQ-51.23\_-|\_\_\_\_\_\_<70|771\_c02c\_0.-1.AQ-35.-16.AAMCaDI\_\_\_\_|\_\_\_\_\_\_<40|771\_cca8\_0.-16.AAMCaDI\_\_\_\_
\caption{The raw fingerprint produced from the active scan.}
\label{fig:Original_fingerprint}
\end{figure}

The source of the domains, along with the label was included in the output file to record the original data source. In all instances the labels mapped to good = 0, bad = 1 and unknown = 3. Each column recorded the deduplicated TLS features across all Client Hellos, giving an accurate representation of the complete server TLS profile. The AS number enrichment was completed by using the pyasn.dat file provided by the original ActiveTLS data set \cite{sosnowski_efactls_2024-1}. 

\subsection{HTTP Header Enrichment}
Unlike the methodology used by Sosnowski et al \cite{sosnowski_efactls_2024-1} whose default relied on just capturing the ‘Server’ header key and value – a commonly overwritten header in CDNs – our approach captures all header keys in order, but disregards the values unless explicitly required and specified in the code. For base usage, only the Server value was kept, all other values were discarded as including them introduced far too much granularity due to the transient nature of many HTTP Headers such as Date, Last-Modified, E-Tag etc. The order of the headers was kept intact as an important metric of mapping similarity between specific applications. 

\begin{table}[htb]
\centering
\scriptsize
\begin{tabular}{cc} 
\hline
fingerprint & http\_header\_hash \\
\hline
71a72d0a2d5478cafb7fc513fe120129a4db5f5dd21671ded5314034b0b72124 & 3898065973 \\
ab545fcff96261433c531d79bd9035d8db4a13b7faef85f5e4283d66ad5ed49d & 3898065973 \\
ab545fcff96261433c531d79bd9035d8db4a13b7faef85f5e4283d66ad5ed49d & 3898065973 \\
71a72d0a2d5478cafb7fc513fe120129a4db5f5dd21671ded5314034b0b72124 & 2350846486 \\
ab545fcff96261433c531d79bd9035d8db4a13b7faef85f5e4283d66ad5ed49d & 1200561793 \\
\end{tabular}
\caption{The TLS fingerprint and the HTTP Header fingerprint output from a Cloudflare scan. The fingerprint is more granular but still contains collisions, indicating that two servers can share the same TLS profile but present different HTTP Headers, enabling differentiation. }
\label{tab:Table_cdn_header_hashes} 
\end{table}

Once collected and correctly processed, the header string was hashed with murmur hash 3. MMH3 is a non-cryptographic hash and selected for its speed, small storage size and good distribution qualities, plus it’s commonly used in open-source tooling such as Shodan for hashing headers and HTML bodies. It is a known quantity for many threat hunters and network analysts and actively used in the cyber security domain.
Table \ref{tab:Table_cdn_header_hashes} shows a sample of the final HTTP header fingerprint alongside the TLS fingerprint for an active scan of domains within the CloudFlare CDN. 

\subsection{Similarly Mapping}
One of the overheads of managing networks and tracking malicious domains with TLS signatures, is the requirement to keep lists of known bad signatures current \cite{papadogiannaki_pump_2023}. This is critical to the methodology remaining useful and viable. Additionally, one of the underlying issues with hash-based fingerprinting is how easy it is to subvert. If a JARM is flagged as being malicious, a threat actor could change a single cipher suite in their server configuration, or change an offered TLS extension, and present an entirely new JARM despite their TLS configuration remaining mostly the same. 

The solution proposed in this paper is to use the produced enriched TLS fingerprint along with similarity mapping aided by visualisation. Despite being complex due to the high dimensionality of the data, similarity comparisons are an effective way of not only eliminating the need for large fingerprint data sets that need to be kept up to date, but they could also prevent malicious actors subverting hash signatures as TLS connections with similar, albeit small changes, would be close in any similarity matrix. 

The biggest barrier to using similarity techniques such as the Jaccard similarity, is the computational overhead required to calculate the comparison matrix of fixed sized vectors when sample sizes increase. Examining TLS traffic across CDNs and network nodes naturally translates to thousands of samples, if not tens of thousands and beyond. Coupled with the requirement of comparing very high dimensional data in the format of multiple TLS extensions, cipher suites, and HTTP headers, the practicality of such similarity mapping becomes unfeasible. 

The adopted approach to address these issues was influenced by the methodology used when mapping chemical fingerprints. Probst and Reymond \cite{probst_visualization_2020} created a library specifically to tackle similarity calculations in very high dimensional datasets called TMAP. TMAP has been used extensively in the chemical science arena to calculate and visualize complex datasets such as the ChEMBL database which contains nodes within the millions. It focuses on nearest neighbour (NN) preservation as a critical feature and avoids linear dimensionality reduction methods such as principal component analysis.

When examining the TLS fingerprint of a server the combination of the various elements can be represented a set, and hence comparisons between servers can be made using Jaccard similarity. Relatively simple, the Jaccard equation \(J(A, B) = |A \cap B| / |A \cup B|\) is a method used to calculate similarity between two sets. Although this is a reasonable approach to take when comparing TLS features, it’s simply unfeasible in practice due to the exhaustive processing require to load each large set before computing the intersection and union of each.  To solve this, the built in TMAP MinHash functionality, in combination with TMAPs local sensitivity hashing was used to create approximate estimations, quickly.

MinHash was specifically chosen because it enables the compression of high dimensional datasets in such a way that the similarity between them can still be deduced \cite{rajaraman_finding_2014}. The technique works by applying multiple hash functions to each set of TLS features to produce a signature for each row. Given a matrix \(M\) of sets, a MinHash function is associated with a number of \(n\) random permutation of the rows.  The hash functions of the permutations can then be described as \(h_1, h_2, h_3 \ldots h_n\) and a vector of \\
\([h_1 (S), h_2 (S), h_3 (S) \ldots h_n (S) ]\) for set \(S\) can be used to construct the MinHash signature. These resultant sets can then be used to form a signature matrix where each column in M is replaced by the MinHash signature for that same column. 

With an approximate method to compress and preserve the similarity, the pairs with the greatest similarity need to be found quickly and effectively. This is solved with a four-phase approach to produce the visualisations. During the first stage the MinHash signatures are index in a local sensitivity hashing forest data structure, enabling \(k-NN\) searches. When initialising the data structure, the same number of hash functions \(d\) is used that was initially used to encode the signatures, along with the number of prefix trees \(l\). A undirected weighted \(c-k-NNG\) graph (\(c\) approximate, \(k\) neighbour) is created from the index points in the LSH forest. This phase takes the argument \(k\) which is the number of nearest neighbours to be searched for. The final phases involve the creation of a minimum spanning tree on the graph using Kruskals algorithm, and the creation of the layout of the tree on the Euclidean plane \cite{probst_visualization_2020}. The approach taken in the four-stage process preserves 1-nearest neighbour relationships well, although the preservation is influenced mainly by the number of hash functions, with the argument of nearest neighbours \(k\) having only minor influences. The benefits of this approach are the speed of the mapping alongside the high locality preservation – which is essential when attempting to map similarity where small vector changes are required to be captured.

\subsection{Fingerprint Processing}
The enriched fingerprint is prepared for similarity mapping and visualisation by creating dynamically sized vectors based upon the totality of the features output from the enrichment process. The raw fingeprint is then interated and each entry appended to a unique entry set, creating a data set of each unique TLS feature and HTTP header, across all servers in the data frame. 

The final stage creates a dictionary of key value pairs based upon these  unique entries. Initially each unique entry is assigned a value of 0 and a second loop through is used to change the value to 1 if that unique entry is seen in the row. This naturally leads to high dimensional vectors that can range from ~80 columns, to ~3000 columns when the HTTP header data is included. 

The final vectors are then used to create the MinHash signature matrix before the LSH forest is generated. To do this, a MinHash object is created and each row in the matrix is encoded. TMAP provides multiple ways to encode the data, including encoding from a weighted array, and string array. In this instance, a pure binary approach was taken as there is nothing inferred from the weight of either the headers, or importantly the TLS features. When focusing on TLS feature sets, the value is in the presence or absence of a specific feature, not the scale of the feature being represented.  

The defining variables for the MinHash to LSH mapping were the arguments used for the number of hash functions, the number of prefix tress \textit{l} and the number of nearest neighbours \textit{k}. For the visualisations produced in this paper, the number of hash functions was defined as 1024, with 128 prefix trees and 100 k nearest neighbours.

\subsection{Visualisations}
\begin{figure}[htb!]
     \centering
     \includegraphics[width=.5\linewidth]{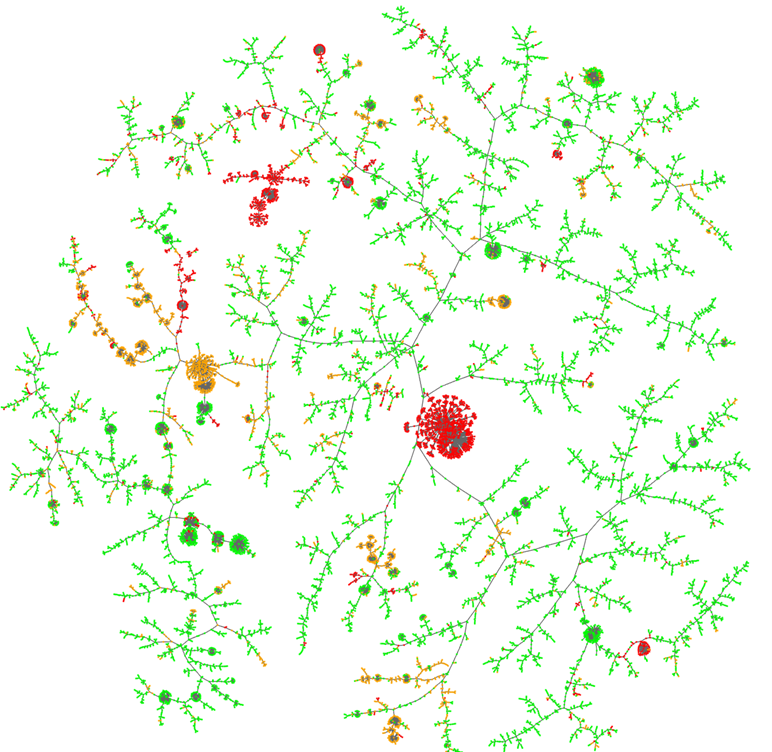}
     \caption{Graph displaying TLS features enriched with HTTP header data. Known good domains are coloured green, known bad red, and unknown orange.}
     \label{fig:fearun_1}
 \end{figure}
The final stage involves initialising and producing the graph. The Python library Fearun was used as it supports millions of data points interactively using web-based visualisations. The TMAP functionality to produce a graph from the LSH index was used to return \textit{x} and \textit{y,} which are the coordinates of the vertices and \textit{s} and \textit{t} which are the ids of the vertices spanning the edges. 

A final plot can be seen in Figure \ref{fig:fearun_1} The dimensionality of the plot is \textit{16254 x 2124} when using TLS features with HTTP header enrichment. 

\section{Evaluation}
Three data sets have been produced during the fingerprint enrichment and similarity experiments. Data set one containing approximately 17711 servers of mixed origin spread across multiple AS. Dataset two that focuses purely on the Cloudflare CDN and contains approximately 5368 servers from the Cloudflare AS 13335, and dataset three, that contains a mixture of 4475 purely malicious domains.

\subsection{Granularity Comparisons}
The fingerprints enriched with HTTP data were noticeably more granular when compared to those generated purely using the TLS data from the ActiveTLS tool. A direct comparison can be made using similar methodology to that used by Sosnowski et al. \cite{sosnowski_active_2023} during their comparison of the fingerprint created with ActiveTLS to those produced using the JARM tool. Unlike the original ActiveTLS stack research, the data sets produced during this research are not broken down by malware classification, as the goal of this research is to identify more broadly the similarity between all types of malicious domains, regardless of their specific classification. The results were encouraging and across all three datasets, all categories saw an increase in granularity. The biggest increases in granularity could be seen in the Cloudflare dataset, where the minimum percentage increase was 66.7\% and the maximum 4523.7\%. The  mixed host data set was the second best performer, with the minimum increase being 4\% and the maximum 199\%. When focusing purely on malicious domains the percentage increases ranged from 4\% to 118\% indicating that there is less diversity in the HTTP Headers.

\subsection{Evaluation of Similarity}
\begin{figure}[!tbp]
  \centering
  \begin{minipage}[b]{0.45\textwidth}
    \includegraphics[width=\textwidth]{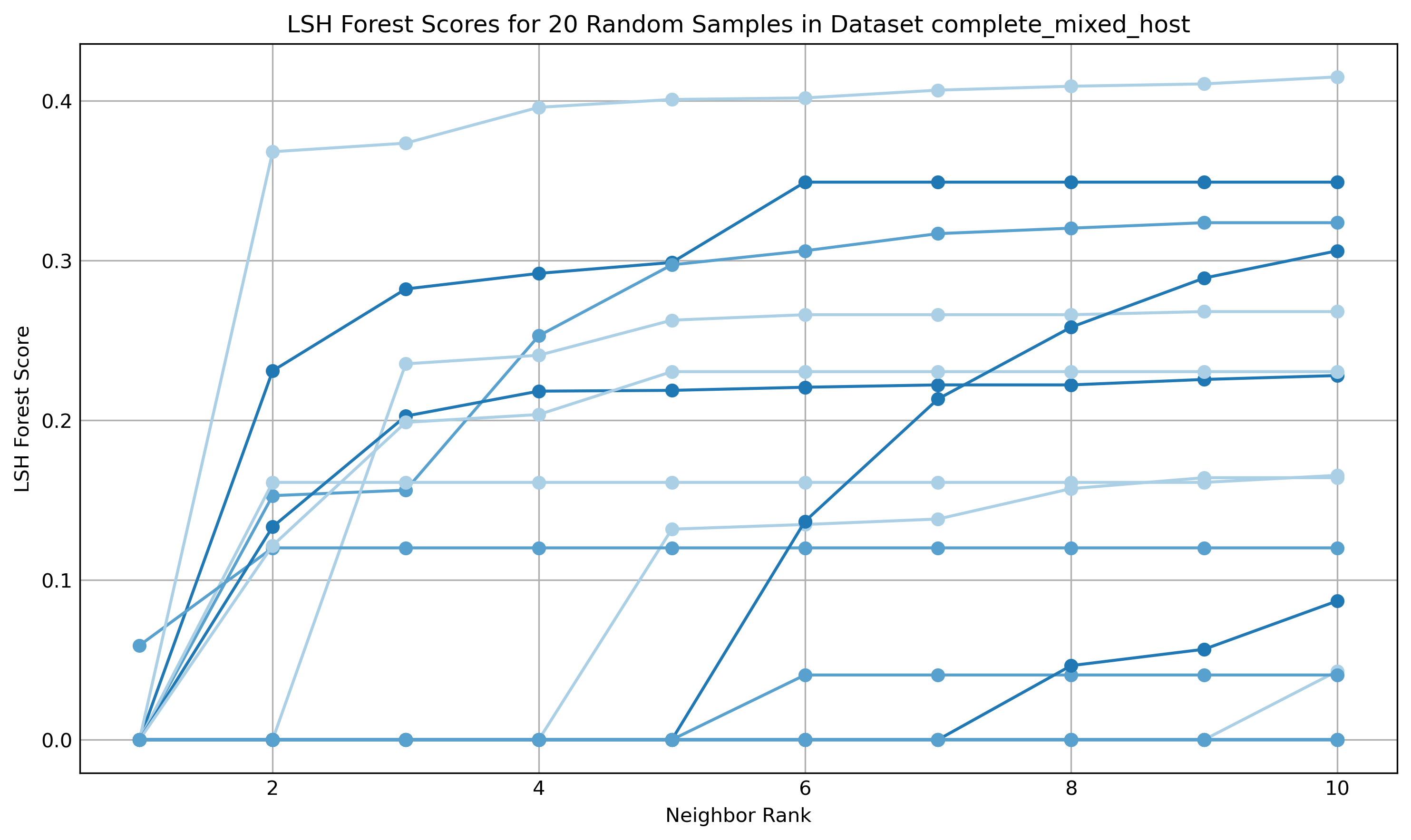}
    \caption{The complete\_mixed\_host dataset displays a diverse number of distance metrics.}
    \label{fig:sim_mixed}
  \end{minipage}
  \hfill
  \begin{minipage}[b]{0.45\textwidth}
    \includegraphics[width=\textwidth]{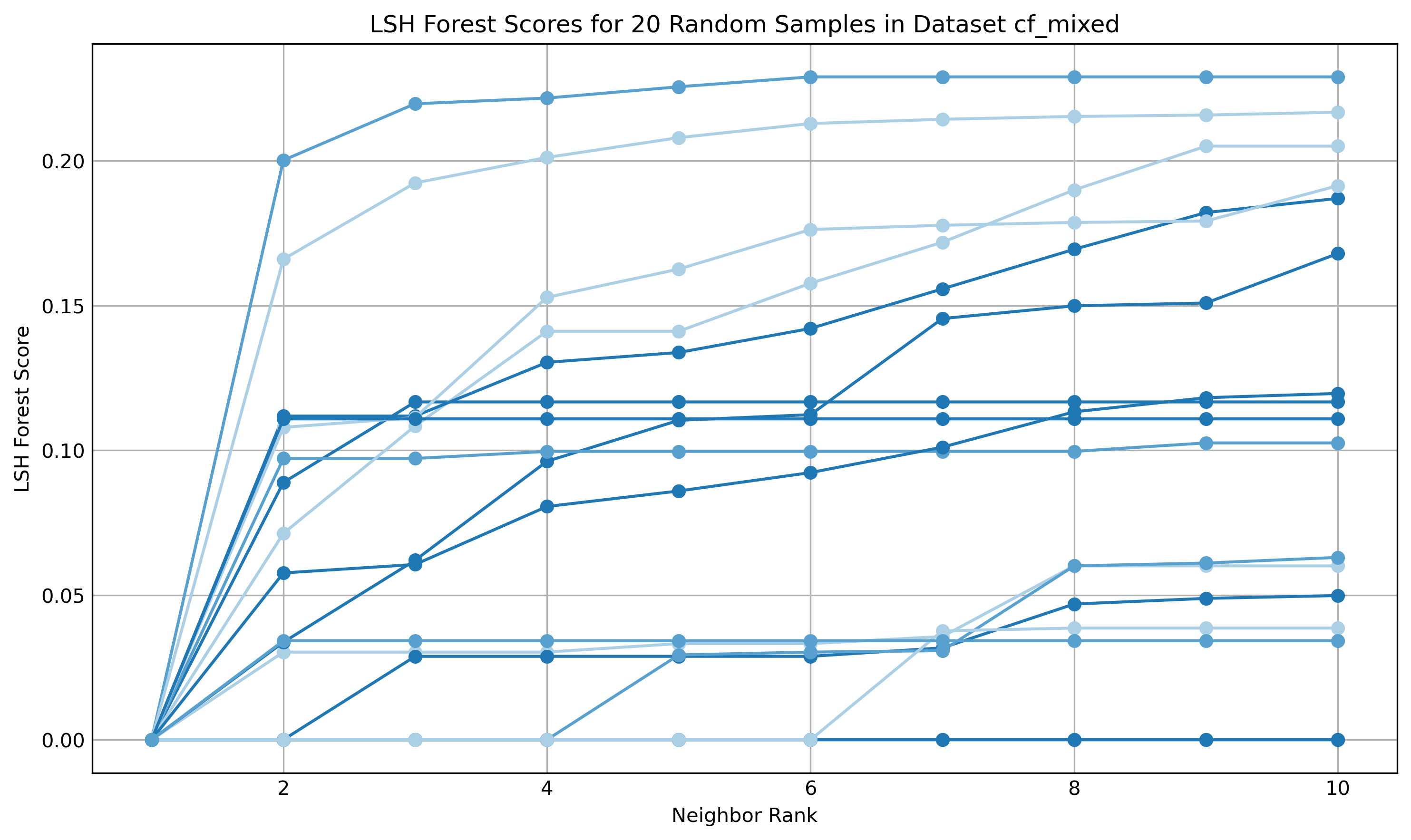}
    \caption{The cf\_mixed dataset displays less diversity in similarity - all 10 k-nearest neighbours are less than 0.30}
    \label{fig:sim_cf}
  \end{minipage}
\end{figure}
One way to determine the effectiveness of the LSH forest produced is to visualise the stability of the similarity calculations. The LSH forest can be queried using a linear scan by id approach which retrieves the k-nearest neighbours to any given id by using a combination of LSH forest traversal and linear scan. This provides an excellent way to visualise and evaluate the success and stability of the distance calculations. In the case of TMAP functionality, the closer the distance to the given ID, the lower the distance value, with an exact match being represented as 0.0. For each data set, 20 random ID’s were queried for their nearest 10 neighbours with the distance from the origin of 0.0 plotted against their k-nearest neighbour. Figure \ref{fig:sim_mixed} shows the result of the complete mixed dataset in comparison to the CloudFlare CDN dataset shown in Figure \ref{fig:sim_cf}. 

\subsection{Visualisation of the LSH Forest}
The graphs produced are visual representations of the \textit{c-k-NNG }construction. In some instances, when the Jaccard distance is 1 when compared to all other data points, the graph can display disconnected nodes and those nodes should be considered outliers. Disconnected graphs can also occur when highly connected clusters of \(\geq k\) occur in the Jaccard space. Although not a clustering method, nodes that share a high similarity should be displayed closely and will form natural clusters with branching. 
Due to the nontraditional approach to clustering it’s difficult to quantify the success or failure of the similarity plotting using techniques such as silhouette scores. Subsequently it has been evaluated by examining distinct areas of the graph to determine if the plots are accurate, consistent, and importantly, if the similarity mapping can determine misclassified or new malicious domains based solely on their similarity to known bad domains. Reclassification's are determined by evaluating the status of the domain in Virus Total, with the status being amended if one or more of the vendors returned a status of anything other than ‘clean’.

\subsubsection{Data Set One}
\begin{figure}[htb!]
    \centering
    \includegraphics[width=1\linewidth]{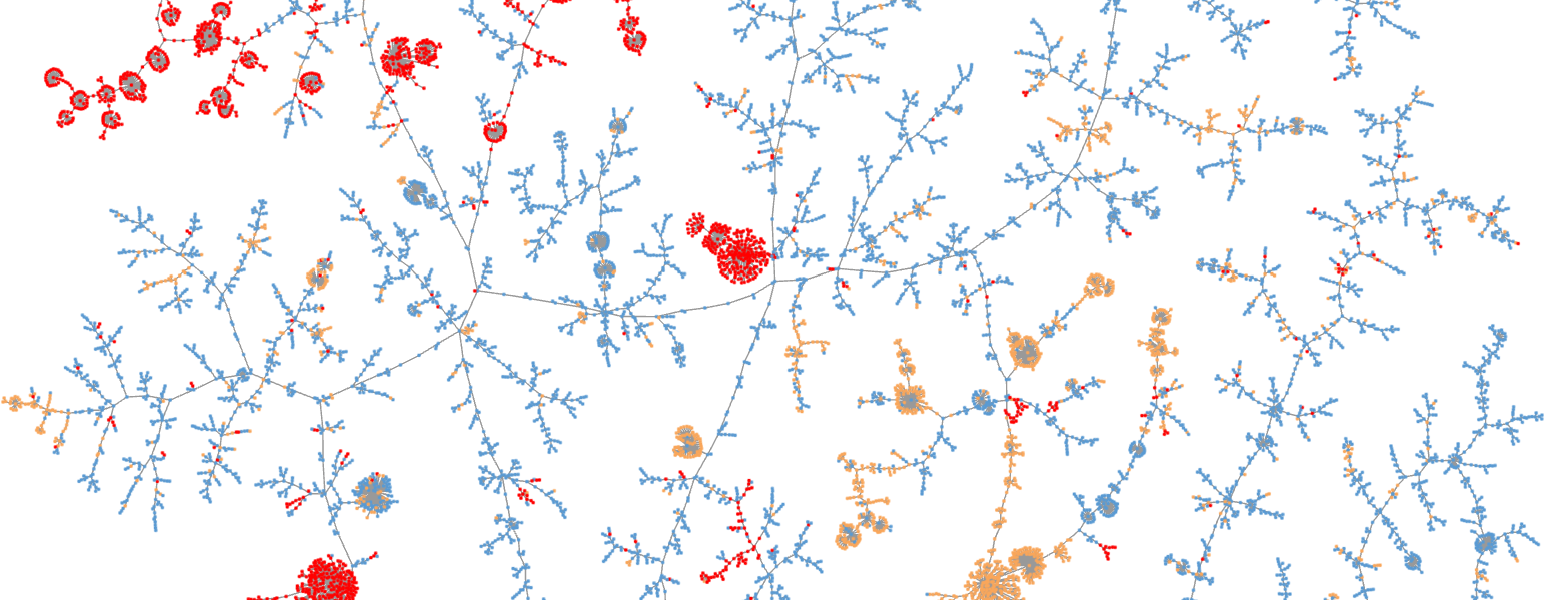}
    \caption{Data set one visualisation. Known bad domains are red, known good are blue and unknown domains are orange.}
    \label{fig:data_set_one_plot}
\end{figure}

For the mixed dataset malicious nodes have formed natural clusters – indicating that there is a high similarity between malware domains across their TLS and HTTP features. Figure \ref{fig:data_set_one_plot} shows a proportion of the final graph where the diversity between known good domains is evident by the high level of branching that lacks clear cluster formation. The dimensionality of the graph for this data set is 17711 x 2133. The statistical analysis of the domains can be found in Table \ref{tab:ds_one}. 
\begin{table}[htb!]
    \centering
    \begin{tabular}{ll}    
    \hline
           Total Nodes: & 211 \\ \hline
           Known Good: & 32 \\ \hline
           Known Bad:& 166\\ \hline
           Unknown: & 13 \\ \hline
           Newly identified malicious domains:& 27\\ \hline
           \multicolumn{2}{l}{Area with 91.47\% confirmed malicious nodes and 12 distinct fingerprints.} \\ 
           \multicolumn{2}{l}{Of the total good or unknown domains 60\% were reclassified to bad.} \\ \hline
           & \\
    \end{tabular}
    \caption{Breakdown of the statistics across three blended areas of the similarity mapping for data set one. }
    \label{tab:ds_one}
\end{table}

\subsubsection{Data Set Two }

\begin{figure}[htb!]
    \centering
    \includegraphics[width=1\linewidth]{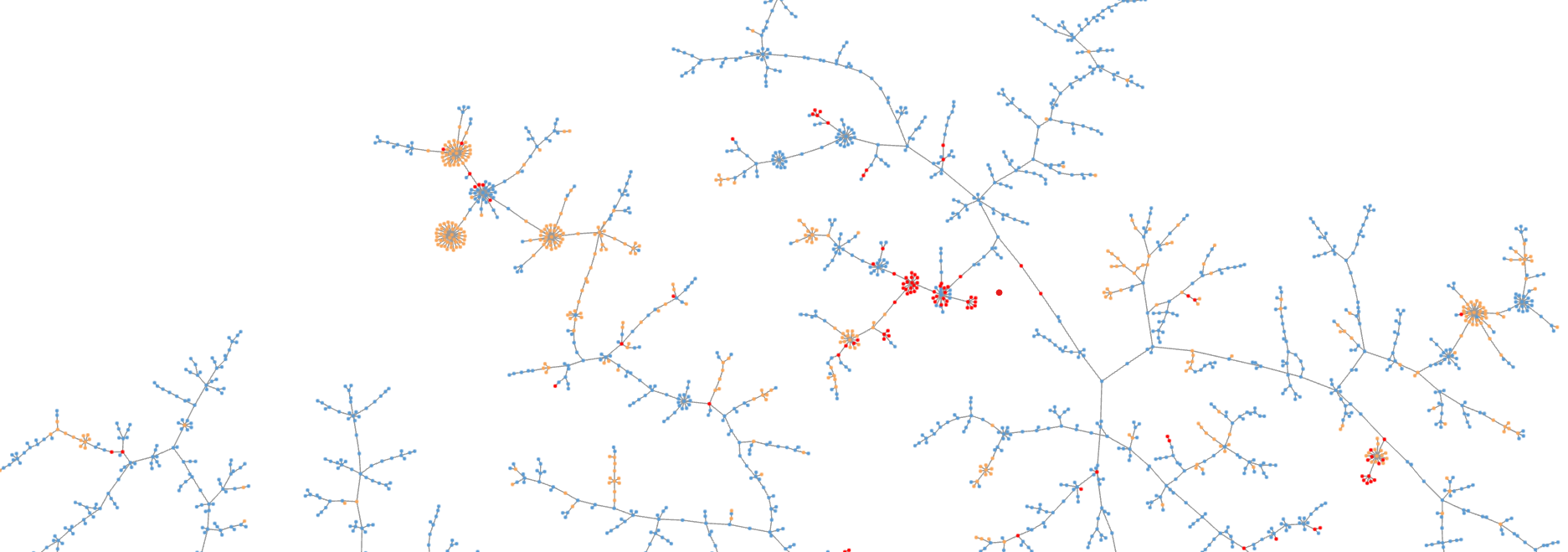}
    \caption{Cloudflare CDN domains visualised with TLS and HTTP header data. Known bad domains are red, known good are blue and unknown domains are orange}
    \label{fig:ds_2_headers}

    \includegraphics[width=1\linewidth]{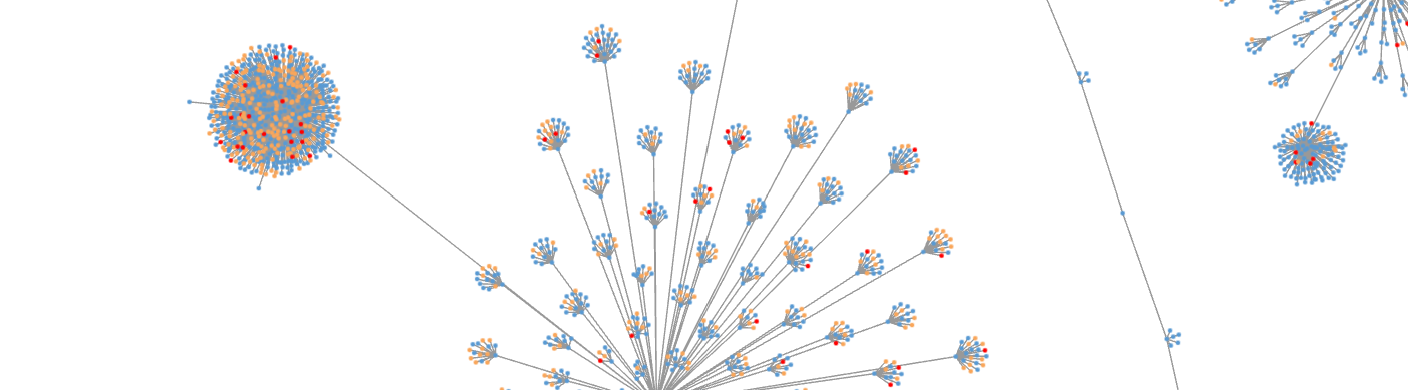}
    \caption{Cloudflare CDN domains visualised with just TLS features, demonstrating a clear lack of diversity in the fingerprints.Known bad domains are red, known good are blue and unknown domains are orange}
    \label{fig:ds_2_no_headers}
\end{figure}
\begin{table}[htb!]
    \centering
    \begin{tabular}{ll}    
    \hline
           Total Nodes: & 85 \\ \hline
           Known Good: & 25 \\ \hline
           Known Bad:& 30 \\ \hline
           Unknown: & 30 \\ \hline
           Newly identified malicious domains:& 40\\ \hline
           \multicolumn{2}{l}{Area with 82.35\% confirmed malicious nodes and 9 distinct fingerprints.} \\ 
           \multicolumn{2}{l}{Of the total good or unknown domains 72.72\% were reclassified to bad.} \\ \hline
           & \\
    \end{tabular}
    \caption{Breakdown of the statistics across three blended areas of the similarity mapping for data set one. }
    \label{tab:ds_two}
\end{table}
Data set two focuses on domains from the Cloudflare AS 13335 and is a blend of known good, known bad and unknown domains. Two graphs have been generated, one that plots the domains solely based on their TLS features, and one based on the full enriched TLS with HTTP features. It was previously established in the results that the diversity of the TLS features across the CDN is poor, with only 59 fingerprints covering the full spectrum of known good domains. This is well represented in the visualisations show in Figure \ref{fig:ds_2_headers} with HTTP headers, compared to Figure \ref{fig:ds_2_no_headers} which is TLS only. The dimensionality of the pure TLS graph is 5368 x 50. In contrast, the dimensionality of the graph when enriched is 5368 x 847.
The statistical analysis of the domains being found in Table \ref{tab:ds_two}.

\subsubsection{Data Set Three }
\begin{figure}[htb!]
    \centering
    \includegraphics[width=1\linewidth]{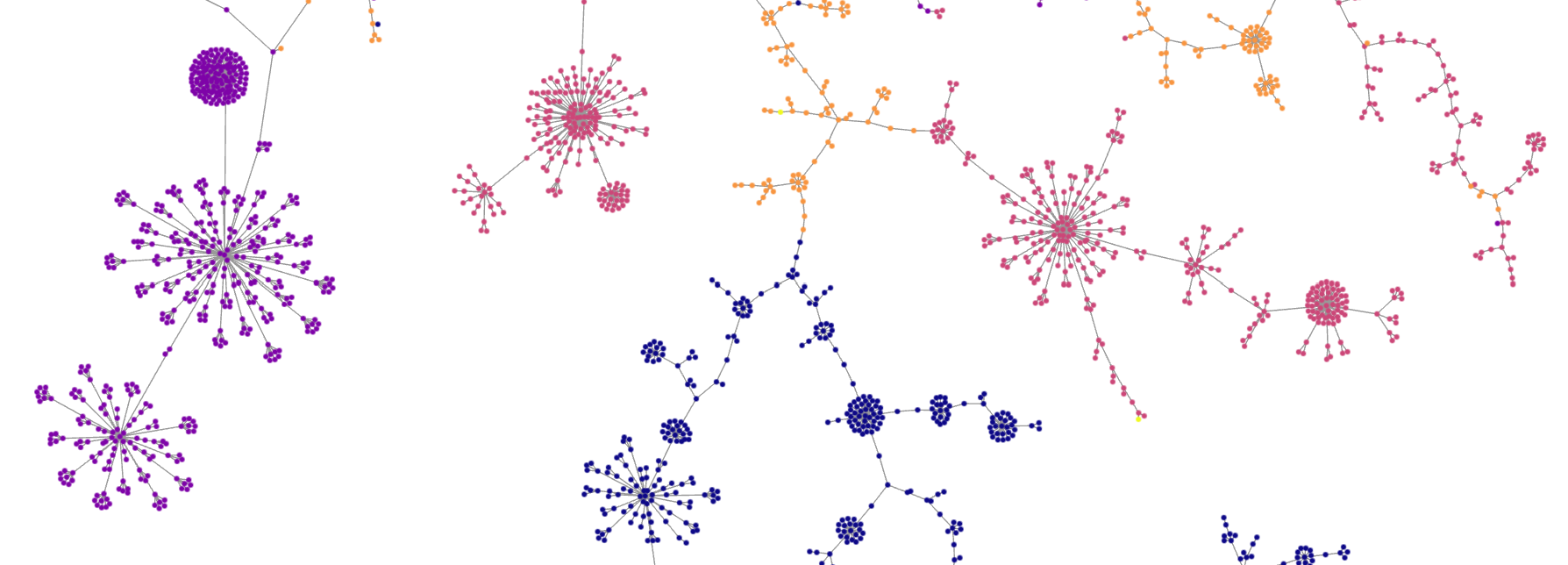}
    \caption{Visualisation of data set three. All domains are malicious and clear clusters can be seen forming by capability. Go Phish domains are seen in yellow, Cert Pl orange, Metasploit pink, Tactical RRM purple and Burp Collaborator Blue.}
    \label{fig:eds_3_plot}
\end{figure}

Data set three represents just domains from five known malicious applications. The use case in this instance is visual confirmation that malicious applications form natural similarity based purely on their external features. Figure \ref{fig:eds_3_plot} displays the final plot based on the enriched TLS with HTTP header features with a dimensionality of 4475 x 306. Clear clusters can be seen forming across applications. Overlap on branches is primarily seen between the Go Phish domains and miscellaneous bad domains provide by Cert.Pl. It’s clear from the plots that despite a range of fingerprints being seen across each of the application, they are close enough in similarity that the applications themselves are close in the Jaccard space.  

\section{Discussion}

The results indicate that a combination of the two-fold approach of increased fingerprint granularity through feature expansion, and similarity mapping, can indeed improve malicious domain detection. Fingerprint granularity was improved in all the sample datasets resulting in improved similarity mapping in each data set with data set two – specifically relating to CDNs being the most improved. As stated in the results section, chapter 4, the aim of the approach was not to create clusters based on features but to calculate and visualize the similarities between application configurations. The fact that many applications form natural clusters is an artifact of the process. It confirms that with many types of malware applications, there are limited variations in the way they are deployed.

One of the major benefits of the approach is its ability to identify evolving threats and hidden malicious domains. The technique of TLS fingerprinting has been used successfully in cyber security for many years now, but the technique is not infallible and small changes to configurations produce very different fingerprints. This leads to a new cycle of detection where new malicious fingerprints must be identified and lists of known malicious fingerprints must be updated. In this situation, it’s difficult to understand changes to malicious applications because the fingerprints themselves are either overly verbose and require analysis to understand the changes, or they are hashed and subsequently cannot be reversed. Although the technique presented here produces a ‘hashed’, enriched fingerprint, the similarity mapping process ensures that small changes are maintained and represented visually. The benefit of this is that new fingerprints that appear close to known branches and clusters of bad domains can be instantly highlighted and flagged for further investigation.

This approach has demonstrated a clear ability to discover new fingerprints and domains that were otherwise unidentified or classified as good. Across data sets one and two, in just the 5 areas of mixed good, bad, and unknown that were evaluated in the results, 67 new malicious domains were discovered based on their similarity to known malicious domains alone. Out of those, 37 were known good domains taken from the Tanco top 1 million \cite{le_pochat_tranco_2019}, indicating that even maintaining and evaluating known goods against the most reputable list is not enough on its own to defend against Phishing and malicious domain

\section{Conclusions}

The article conducts a critical literature review evaluating active scanning techniques that can be used to generate server fingerprints. Following the review and having identified the pros and cons of existing methods the article presents the design and development of an active scanning fingerprint. The proposed method increases the granularity of current techniques, improving the ability to detect malicious domains hosted on CDNs where feature similarity is high. Moreover the article, enhances the applicability and security of fingerprinting by introducing a suitable similarity mapping approach. Finally we critically evaluate the results and findings from practical experiments. It’s ability to reclassify known good domains from high reputational allow lists, even when examining only small subsections of the c-k-NNG graph demonstrates there is future potential in the technique





\bibliographystyle{elsarticle-num} 
\bibliography{ref.bib}

\begin{thebibliography}{10}
\expandafter\ifx\csname url\endcsname\relax
  \def\url#1{\texttt{#1}}\fi
\expandafter\ifx\csname urlprefix\endcsname\relax\def\urlprefix{URL }\fi
\expandafter\ifx\csname href\endcsname\relax
  \def\href#1#2{#2} \def\path#1{#1}\fi

\bibitem{rescorla2018transport}
E.~Rescorla, The transport layer security (tls) protocol version 1.3, Tech. rep. (2018).

\bibitem{warburton20212021}
D.~Warburton, S.~Vinberg, The 2021 tls telemetry report, F5 Labs. WWW-dokumentti. Saatavissa: https://www. f5. com/labs/articles/threat-intelligence/the-2021-tls-telemetry-report [viitattu 11.2. 2023] (2021).

\bibitem{oh_survey_2022}
C.~Oh, J.~Ha, H.~Roh, A {Survey} on {TLS}-{Encrypted} {Malware} {Network} {Traffic} {Analysis} {Applicable} to {Security} {Operations} {Centers}, Applied Sciences (2022).

\bibitem{shamsimukhametov_is_2022}
D.~Shamsimukhametov, A.~Kurapov, M.~Liubogoshchev, E.~Khorov, Is {Encrypted} {ClientHello} a {Challenge} for {Traffic} {Classification}?, IEEE Access (2022).

\bibitem{sosnowski_active_2023}
M.~Sosnowski, J.~Zirngibl, P.~Sattler, G.~Carle, C.~Grohnfeldt, M.~Russo, D.~Sgandurra, Active {TLS} {Stack} {Fingerprinting}: {Characterizing} {TLS} {Server} {Deployments} at {Scale}, arXiv:2206.13230 [cs] (Aug. 2023).

\bibitem{lindeman_easily_2020}
J.~A. Lindeman, Laura, Easily {Identify} {Malicious} {Servers} on the {Internet} with {JARM} (Nov. 2020).

\bibitem{papadogiannaki_pump_2023}
E.~Papadogiannaki, S.~Ioannidis, Pump {Up} the {JARM}: {Studying} the {Evolution} of {Botnets} {Using} {Active} {TLS} {Fingerprinting}, in: 2023 {IEEE} {Symposium} on {Computers} and {Communications} ({ISCC}), 2023, pp. 764--770.

\bibitem{matousek_reliability_2021}
P.~Matoušek, I.~Burgetová, O.~Ryšavý, M.~Victor, On {Reliability} of {JA3} {Hashes} for {Fingerprinting} {Mobile} {Applications}, in: S.~Goel, P.~Gladyshev, D.~Johnson, M.~Pourzandi, S.~Majumdar (Eds.), Digital {Forensics} and {Cyber} {Crime}, Lecture {Notes} of the {Institute} for {Computer} {Sciences}, {Social} {Informatics} and {Telecommunications} {Engineering}, Springer International Publishing, Cham, 2021, pp. 1--22.

\bibitem{van_der_mandele_encrypted_2023}
A.~Van Der~Mandele, A.~Ghendini, C.~Wood, R.~Mehra, \href{https://blog.cloudflare.com/announcing-encrypted-client-hello}{Encrypted {Client} {Hello} - the last puzzle piece to privacy} (Sep. 2023).
\newline\urlprefix\url{https://blog.cloudflare.com/announcing-encrypted-client-hello}

\bibitem{bhandari_2022_2022}
H.~Bhandari, J.~Viggiano, \href{https://almanac.httparchive.org/en/2022/cdn}{The 2022 {Web} {Almanac}: {CDN}}, Tech. rep., HTTP Archive, issue: 22 Publication Title: The 2022 Web Almanac Volume: 4 (Oct. 2022).
\newline\urlprefix\url{https://almanac.httparchive.org/en/2022/cdn}

\bibitem{siby_evaluating_2023}
S.~Siby, L.~Barman, C.~Wood, M.~Fayed, N.~Sullivan, C.~Troncoso, \href{https://petsymposium.org/popets/2023/popets-2023-0099.php}{Evaluating practical {QUIC} website fingerprinting defenses for the masses}, Proceedings on Privacy Enhancing Technologies (2023).
\newline\urlprefix\url{https://petsymposium.org/popets/2023/popets-2023-0099.php}

\bibitem{tang_hslf_2021}
Z.~Tang, Q.~Wang, W.~Li, H.~Bao, F.~Liu, W.~Wang, {HSLF}: {HTTP} {Header} {Sequence} {Based} {LSH} {Fingerprints} for {Application} {Traffic} {Classification}, in: Computational {Science} – {ICCS} 2021, Springer International Publishing, 2021, pp. 41--54.

\bibitem{mcgahagan_comprehensive_2019}
J.~McGahagan, D.~Bhansali, M.~Gratian, M.~Cukier, A {Comprehensive} {Evaluation} of {HTTP} {Header} {Features} for {Detecting} {Malicious} {Websites}, in: 2019 15th {European} {Dependable} {Computing} {Conference} ({EDCC}), IEEE, Naples, Italy, 2019, pp. 75--82.

\bibitem{al-hakimi_hunting_2021}
S.~Al-Hakimi, F.~Bax, Hunting for malicious infrastructure using big data, Tech. rep., University of Amsterdam (Feb. 2021).

\bibitem{bortolameotti_headprint_2020}
R.~Bortolameotti, T.~van Ede, A.~Continella, T.~Hupperich, M.~H. Everts, R.~Rafati, W.~Jonker, P.~Hartel, A.~Peter, {HeadPrint}: detecting anomalous communications through header-based application fingerprinting, in: Proceedings of the 35th {Annual} {ACM} {Symposium} on {Applied} {Computing}, {SAC} '20, Association for Computing Machinery, New York, NY, USA, 2020, pp. 1696--1705.

\bibitem{sosnowski_efactls_2024-1}
M.~Sosnowski, J.~Zirngibl, P.~Sattler, G.~Carle, C.~Grohnfeldt, M.~Russo, D.~Sgandurra, {EFACTLS}: {Effective} {Active} {TLS} {Fingerprinting} for {Large}-scale {Server} {Deployment} {Characterization}, IEEE Transactions on Network and Service Management (2024) 1--1Conference Name: IEEE Transactions on Network and Service Management.
\newblock \href {https://doi.org/10.1109/TNSM.2024.3364526} {\path{doi:10.1109/TNSM.2024.3364526}}.

\bibitem{le_pochat_tranco_2019}
V.~Le~Pochat, T.~Van~Goethem, S.~Tajalizadehkhoob, M.~Korczynski, W.~Joosen, Tranco: {A} {Research}-{Oriented} {Top} {Sites} {Ranking} {Hardened} {Against} {Manipulation}, in: Proceedings 2019 {Network} and {Distributed} {System} {Security} {Symposium}, Internet Society, San Diego, CA, 2019.

\bibitem{tilborghs_flagging_2022}
S.~Tilborghs, Flagging 13 {Million} {Malicious} {Domains} in 1 {Month} with {Newly} {Observed} {Domains} (Sep. 2022).

\bibitem{Blechschmidt}
Blechschmidt, \href{https://github.com/blechschmidt/massdns}{Blechschmidt/massdns: A high-performance dns stub resolver for bulk lookups and reconnaissance (subdomain enumeration)}.
\newline\urlprefix\url{https://github.com/blechschmidt/massdns}

\bibitem{probst_visualization_2020}
D.~Probst, J.-L. Reymond, Visualization of very large high-dimensional data sets as minimum spanning trees, Journal of Cheminformatics 12~(1) (2020) 12.
\newblock \href {https://doi.org/10.1186/s13321-020-0416-x} {\path{doi:10.1186/s13321-020-0416-x}}.

\bibitem{rajaraman_finding_2014}
Finding {Similar} {Items}, in: A.~Rajaraman, J.~D. Ullman, J.~Leskovec (Eds.), Mining of {Massive} {Datasets}, 2nd Edition, Cambridge University Press, Cambridge, 2014, pp. 68--122.
\newblock \href {https://doi.org/10.1017/CBO9781139924801.004} {\path{doi:10.1017/CBO9781139924801.004}}.

\end{thebibliography}

\end{document}